%% file: 3PNSOPaperRapid.tex
\documentclass[a4paper,twoside]{article}

\usepackage{ifthen}
\newboolean{adp}
\setboolean{adp}{false}
\newboolean{arxiv}
\setboolean{arxiv}{true}
\newboolean{notadp}
\setboolean{notadp}{true}
\ifadp
\setboolean{notadp}{false}
\fi
\newboolean{notarxiv}
\setboolean{notarxiv}{true}
\ifarxiv
\setboolean{notarxiv}{false}
\fi

\usepackage{times}

\usepackage[american]{babel}

\ifnotadp
\usepackage{a4wide}
\input{fleqn.clo}
\PassOptionsToPackage{fleqn}{amsmath}
\PassOptionsToPackage{fleqn}{amstex}
\fi

\usepackage[fleqn]{amsmath}
\usepackage{amsfonts}
\usepackage{amssymb}

\allowdisplaybreaks
\ifnotarxiv
\usepackage{xcolor}
\xdefinecolor{mylinkcolor}{rgb}{0,0,0.5}
\usepackage[
	bookmarksnumbered, bookmarksopen, bookmarksopenlevel=2,
	breaklinks=true, colorlinks=true,
	filecolor=mylinkcolor, citecolor=mylinkcolor, linkcolor=mylinkcolor,
	urlcolor=mylinkcolor, menucolor=mylinkcolor,
]{hyperref}
\fi

\ifarxiv % breaklinks must be removed
\usepackage{xcolor}
\xdefinecolor{mylinkcolor}{rgb}{0,0,0.5}
\usepackage[
	bookmarksnumbered, bookmarksopen, bookmarksopenlevel=1,
	colorlinks=true,
	filecolor=mylinkcolor, citecolor=mylinkcolor, linkcolor=mylinkcolor,
	urlcolor=mylinkcolor, menucolor=mylinkcolor,
]{hyperref}
\fi

\ifnotadp
\usepackage[merge,square,comma,numbers,sort&compress]{natbib}
\fi

\input{macros.tex}

\newcommand{\mycite}[2]{% \mycite{without merging (AdP), with merging (non AdP}
\ifnotadp\cite{#2}\fi%
\ifadp\cite{#1}\fi}

\begin{document}

\def\tit{Next-to-next-to-leading order post-Newtonian spin-orbit Hamiltonian for self-gravitating binaries}
\def\titleShort{NNLO spin-orbit Hamiltonian for self-gravitating binaries}

\def\Fauthor{Johannes Hartung}
\def\FauthorShort{J.\ Hartung}
\def\Sauthor{Jan Steinhoff}
\def\SauthorShort{J.\ Steinhoff}

\def\FauthorFootnote{Corresponding author\quad
	E-mail:~\textsf{johannes.hartung@uni-jena.de},
	Phone: +49\,3641\,947\,106,
	Fax: +49\,3641\,947\,102}

\def\adr{Theoretisch--Physikalisches Institut, \\
	Friedrich--Schiller--Universit\"at, \\
	Max--Wien--Platz 1, 07743 Jena, Germany, EU}

\def\abs{We present the next-to-next-to-leading order post-Newtonian (PN)
spin-orbit Hamiltonian for two self-gravitating spinning compact objects. If at
least one of the objects is rapidly rotating, then the corresponding interaction
is comparable in strength to a 3.5PN effect. The result in the present paper in
fact completes the knowledge of the post-Newtonian Hamiltonian for binary
spinning black holes up to and including 3.5PN. The Hamiltonian is checked via
known results for the test-spin case and via the global Poincar\'e algebra with
the center-of-mass vector uniquely determined by an ansatz.}

\def\pacs{04.25.Nx, 04.20.Fy, 04.25.-g, 97.80.-d, 45.50.Jf}
\def\keyw{post-Newtonian approximation; canonical formalism;
	approximation methods; equations of motion; binary stars}

\def\ack{
We thank G. Sch\"afer for suggesting this interesting research project, for very useful
and encouraging discussions, and for strongly supporting this work.
We also thank P.\ Jaranowski for sharing his insight in the calculation of
the 3PN point-mass Hamiltonian and for providing several test integrals, 
and T.\ Damour for very useful comments on our manuscript regarding the $\dim$-dimensional 
UV-analysis.
We further gratefully acknowledge many useful discussions 
with D.\ Brizuela about {\scshape xPert} and doing perturbation theory in 
arbitrary dimensions, 
with M.\ Q.\ Huber about three-body integral related Appell $F_4$ functions,
with S.\ Hergt about the Poincar\'e algebra and center-of-mass vector ans\"atze, 
with M.\ Tessmer about UV-analysis optimization issues,
and with J.\ Sperrhake about the manuscript and simplifications of the angular averagings in prolate spheroidal 
coordinates. 
JS further thanks A. Buonanno for useful discussions on the
canonical formalism in the test-spin case.
This work is funded by the Deutsche Forschungsgemeinschaft (DFG) through
the Research Training Group GRK 1523 ``Quanten- und Gravitationsfelder'' and
the Collaborative Research Center SFB/TR7 ``Gravitations\-wellen\-astronomie.''
}

\ifnotadp
\title{\tit}
\author{\Fauthor\thanks{\FauthorFootnote}, \quad \Sauthor%, \quad \Tauthor
	\bigskip \\ \adr}
\maketitle
\begin{abstract}
\abs
\end{abstract}
\noindent PACS numbers: \pacs \\ Keywords: \keyw
\fi

\ifadp
\title[\titleShort]{\tit}
\author[\FauthorShort]{\Fauthor\footnote{\FauthorFootnote}}
\author[\SauthorShort]{\Sauthor}
%\author[\TauthorShort]{\Tauthor}
\address{\adr}
\keywords{\keyw}
\begin{abstract}
\abs
\end{abstract}
\maketitle
\fi

\ifnotarxiv
\hypersetup{pdftitle={\tit}, pdfauthor={\Fauthor, \Sauthor}}
\fi

%
% document starts here
%

\section{Introduction}\label{sec:intro}
In the present paper the next-to-next-to-leading order post-Newtonian (PN)
spin-orbit Hamiltonian for two self-gravitating spinning compact objects is
derived. This Hamiltonian is of the order 3.5PN if at least one of the objects
is rapidly rotating. Indeed, the result in the present paper completes the
knowledge of the post-Newtonian Hamiltonian (and thus of the equations of
motion) for binary spinning black holes up to and including 3.5PN. Besides the
well-known Newtonian and 1PN Hamiltonians, previous results to this order for
point-masses are the 2PN \cite{Ohta:Okamura:Kimura:Hiida:1974,
Damour:Schafer:1985, Damour:Schafer:1988}, 2.5PN \cite{Schafer:1985,
Schafer:1986}, 3PN \cite{Jaranowski:Schafer:1998, Kimura:Toiya:1972,
Jaranowski:Schafer:1999, Damour:Jaranowski:Schafer:2000,
Damour:Jaranowski:Schafer:2001}, and 3.5PN \cite{Jaranowski:Schafer:1997,
Konigsdorffer:Faye:Schafer:2003} Hamiltonians. For the spin part the leading
order can be found in \cite{Barker:OConnell:1975, DEath:1975,
Barker:OConnell:1979, Thorne:Hartle:1985} and the next-to-leading order in
\cite{Damour:Jaranowski:Schafer:2008:1, Steinhoff:Hergt:Schafer:2008:2,
Steinhoff:Hergt:Schafer:2008:1}. At the 3.5PN level one also
needs all Hamiltonians cubic in the spins derived in \cite{Hergt:Schafer:2008:2,
Hergt:Schafer:2008}. Notice that so far these cubic Hamiltonians are known for
binary black holes only, whereas all other mentioned Hamiltonians (including the
one derived in the present paper) are valid for general compact objects [or have
been generalized to this case, see \mycite{Poisson:1998, Porto:Rothstein:2008:2, 
Steinhoff:Schafer:2009:1, Hergt:Steinhoff:Schafer:2010:1}
{Poisson:1998, Porto:Rothstein:2008:2,
*Porto:Rothstein:2008:2:err, Steinhoff:Schafer:2009:1,
Hergt:Steinhoff:Schafer:2010:1} for the spin(1)-spin(1) level]. Further, tidal
effects become very important for general compact objects like neutron stars,
see, e.g., \cite{Damour:Nagar:2009} and also \cite{Vines:Flanagan:2010}.

The calculation in the present paper was performed within the canonical
formalism of Arnowitt, Deser, and Misner (ADM) \cite{Arnowitt:Deser:Misner:1962,
Arnowitt:Deser:Misner:2008}, which was recently generalized to rotating objects
at linear order in spin \cite{Steinhoff:Schafer:2009:2}, see also
\cite{Steinhoff:2011, Steinhoff:Wang:2009, Steinhoff:Schafer:Hergt:2008}. For
various other (noncanonical) derivations of post-Newtonian results at the
point-mass level see \cite{Blanchet:2006, Futamase:Itoh:2007, Pati:Will:2000,
Goldberger:Rothstein:2006, Gilmore:Ross:2008,
Kol:Smolkin:2009, Foffa:Sturani:2011} and references therein. The
next-to-leading order in spin was also treated in
{\mycite{Tagoshi:Ohashi:Owen:2001, Faye:Blanchet:Buonanno:2006, Perrodin:2010,
Porto:2010, Levi:2010, Porto:Rothstein:2008:1, Levi:2008}
{Tagoshi:Ohashi:Owen:2001, Faye:Blanchet:Buonanno:2006, Perrodin:2010,
Porto:2010, Levi:2010, Porto:Rothstein:2008:1, *Porto:Rothstein:2008:1:err,
Levi:2008}}. Also more than two compact objects have been treated at high
post-Newtonian orders \cite{Schafer:1987, Ohta:Kimura:Hiida:1975,
Lousto:Nakano:2008, Chu:2009, Hartung:Steinhoff:2010}. Notice that the
calculation in the present paper is comparable in complexity to the one of the
3PN point-mass Hamiltonian. In particular one has to check for certain integral
contributions that can only be handled correctly within dimensional
regularization, as they may lead to ambiguities when integrated in three spatial
dimensions \cite{Damour:Jaranowski:Schafer:2001,
Damour:Jaranowski:Schafer:2008:2}. The result of the present paper was verified
using the test-spin Hamiltonian in Kerr-spacetime given in
\cite{Barausse:Racine:Buonanno:2009} and using the global Poincar\'e algebra.
For the latter the center-of-mass vector had to be determined uniquely from an
ansatz with 68 coefficients.

It should be noted that the Hamiltonian derived in the present paper is yet only
useful within the Taylor-expanded post-Newtonian series if at least one of the
objects is rapidly rotating (due to the missing 4PN point-mass Hamiltonian).
However, if the 4PN point-mass Hamiltonian can be derived in the future, then
the result of the present paper must be included in the post-Newtonian series
also when the spins are small. But the effective one-body approach for
nonspinning objects is able to cover such higher post-Newtonian orders by
calibration to numerical relativity, see \cite{Damour:Nagar:2009:2,
Buonanno:etal:2009} and references therein. The result given in the present
paper is thus expected to be valuable for the effective one-body formalism, just
like the next-to-leading order one \cite{Damour:Jaranowski:Schafer:2008:3,
Pan:etal:2009, Barausse:Buonanno:2009}.

The paper is organized as follows. In Sect.\ \ref{sec:outline} a brief outline
of the calculation is given. The next-to-next-to-leading order spin-orbit
Hamiltonian is presented in Sect.\ \ref{sec:result}. A comparison with known
results for the test-spin case is made. The Hamiltonian is further checked via
the global Poincar\'e algebra in Sect.\ \ref{sec:poincare}, where the
center-of-mass vector is uniquely determined from an ansatz. In forthcoming
papers we will derive the next-to-next-to-leading order spin(1)-spin(2)
Hamiltonian and provide much more details on the calculation of the spin-orbit
Hamiltonian, shown in the present paper, as well.

Three-dimensional vectors are written in boldface and their components are
denoted by Latin indices. The scalar product between two vectors $\vct{a}$ and 
$\vct{b}$ is denoted by $(\vct{a}\vct{b}) \equiv (\vct{a} \cdot \vct{b})$.
Our units are such that $c=1$. There is no special convention for Newton's gravitational
constant $\gravthree$. In the results $\vmom{a}$ denotes the canonical linear momentum of the $a$th 
object, $\hat{\vct{z}}_a$ the position of the object, $m_a$ the mass of the object, 
$\hat{\vct{S}}_a$ the spin of the object, $r_{ab}=|\hat{\vct{z}}_a - \hat{\vct{z}}_b|$ the 
relative distance between two objects, and $\vnxa{ab} = (\hat{\vct{z}}_a - \hat{\vct{z}}_b)/r_{ab}$ 
the direction vector pointing from object $b$ to object $a$.
In the integrands $r_a = |\vct{x} - \hat{\vct{z}}_a|$, $\vnxa{a} = (\vct{x}-\hat{\vct{z}}_a)/r_a$,
and $s_{ab} = r_a + r_b + r_{ab}$. In the binary case the object labels $a, b$ take only the values $1$ and $2$.

\section{Outline of the calculation}\label{sec:outline}
In the following we present a short outline of our calculation -- which will be discussed more detailed
in a forthcoming publication -- and cite the main literature necessary to undertake it.

For all computations we used {\scshape xTensor} \cite{MartinGarcia:2002}, a free package
for {\scshape Mathematica} \cite{Wolfram:2003}, especially because of its fast index
canonicalizer based on the package {\scshape xPerm} \cite{MartinGarcia:2008}. We also used
the package {\scshape xPert} \cite{Brizuela:MartinGarcia:MenaMarugan:2009}, which is part of {\scshape xTensor},
for performing the perturbative part of our calculations. Furthermore we wrote several 
{\scshape Mathematica} packages ourselves for evaluating integrals.

First we generalized the derivation of the canonical formalism given in \cite{Steinhoff:Schafer:2009:2} to arbitrary
dimensions. The initial action is of the same form as in $\dim=3$ and the $(\dim+1)$-split is also straightforward. The $\dim$-dependence
enters the calculation via the relation between extrinsic curvature and field momentum, and via the decomposition of the metric and field momentum
in the ADM transverse-traceless gauge.
The calculation was done in $\dim$ dimensions
due to possible appearance of ambiguities in three-dimensional integrals. Ambiguity means that one will get
different results when one does an integration by parts in a certain integral. These ambiguities can 
only be ruled out or corrected by doing the UV-analysis explained in 
\cite{Damour:Jaranowski:Schafer:2001, Damour:Jaranowski:Schafer:2008:2}, which
relies on the complete $\dim$-dependence of the integrands. In the following UV-analysis always refers 
to the analysis of certain integrals via a Taylor expansion of the field expressions in the $r_1$ variable, 
extracting the critical $r_1$ powers, and averaging over the $\vnxa{1}$ vectors afterwards to get the pole part 
of the integrals in $\varepsilon \equiv \dim-3$.

The integrations by parts necessary to get the Hamiltonian presented in this paper were done like 
suggested in \cite{Jaranowski:Schafer:1998,Damour:Jaranowski:Schafer:2001,Steinhoff:Wang:2009} to get 
comparable intermediate results. 
After accomplishing the integration by parts we ended up with a Hamiltonian density which can be split up into 
three parts: A kinetic field part (containing the kinetic energy of the propagating field degrees of freedom) 
a matter part (containing only field-matter-interactions), and an interaction part (containing interactions 
between matter fields and the propagating fields), see, e.g., \cite{Schafer:1985, Schafer:1986}. We checked intermediate
results against \cite{Jaranowski:Schafer:1998,Steinhoff:Schafer:Hergt:2008,Steinhoff:Schafer:2009:2,Steinhoff:Wang:2009}.

From the Hamiltonian obtained by integrating the density from above one can go to a Routhian (a 
Hamiltonian in matter degrees of freedom and a Lagrangian in the propagating field degrees of freedom) 
as suggested in \cite{Jaranowski:Schafer:1998,Steinhoff:Schafer:Hergt:2008} and can eliminate 
the propagating degrees of freedom by inserting their approximate solutions in terms of the matter variables
\cite{Jaranowski:Schafer:1998}. Subsequent elimination of time derivatives
corresponds to a coordinate transformation \cite{Schafer:1984, Damour:Schafer:1991}.

After obtaining a suitable Hamiltonian density via the simplifications mentioned in the last paragraphs,
one has to integrate all appearing terms in the density to get a full Hamiltonian. The appearing integrals 
can be divided into three types: 
the delta-type $\int {\rm d}^\dim x f(\vct{x}) \dl{1}$, 
the Riesz-type $\int {\rm d}^\dim x\, n^{i_1}_1 \dots n^{i_k}_1 n^{j_1}_2 \dots n^{j_\ell}_2 r_1^\alpha r_2^\beta$
and the generalized Riesz-type 
$\int {\rm d}^3 x\, n^{i_1}_1 \dots n^{i_k}_1 n^{j_1}_2 \dots n^{j_\ell}_2 r_1^\alpha r_2^\beta s_{12}^\gamma$.

The delta-type integrals can be solved by the Partie-Finie regularization procedure 
%\remark{in $\dim=3$}
mentioned in the appendices of, e.g., 
\cite{Jaranowski:Schafer:1997,Jaranowski:Schafer:1998,Steinhoff:Schafer:Hergt:2008}.
%\remark{In arbitrary dimensions the delta-type integrals are always finite for a dense set of 
% values of the dimension $\dim$ and
% therefore the regularized value follows from analytic continuation.}
Another possibility to solve some of them is via the Riesz kernel method, where one inserts a Riesz kernel for the
delta functions \mycite{Riesz:1949,Damour:Jaranowski:Schafer:2008:2}{Riesz:1949,*Riesz:1949:err,Damour:Jaranowski:Schafer:2008:2}. 
This was done as an alternative way to check whether our algorithms work correctly,
although we did not calculate all delta-type integrals via the Riesz kernel method because not all appearing 
inverse Laplacians can be solved via the method mentioned in \cite{Jaranowski:Schafer:1997}.
Of course, using the Riesz kernel instead of a delta as source of the fields (to eliminate the necessity of distributional 
derivatives) makes the integration much more complicated because all 
delta-type integrals will be changed into integrals of the Riesz-type or the generalized Riesz-type.

The Riesz-type integrals can be solved by eliminating the $\vnxa{1}$ and $\vnxa{2}$ vectors via
rewriting them into derivatives as shown in \cite{Jaranowski:Schafer:1998,Steinhoff:Schafer:Hergt:2008} and
solving the remaining scalar integrals via the Riesz-formula \mycite{Riesz:1949,Damour:Jaranowski:Schafer:2008:2}
{Riesz:1949,*Riesz:1949:err,Damour:Jaranowski:Schafer:2008:2}
\begin{align}
 \left.\int {\rm d}^\dim x\, r_1^\alpha r_2^\beta\right|_{\text{reg}} & = 
\pi^{\dim/2} \frac{
	\Gamma\left(\frac{\alpha+\dim}{2}\right)\Gamma\left(\frac{\beta+\dim}{2}\right)\Gamma\left(-\frac{\alpha+\beta+\dim}{2}\right)
	}{
	\Gamma\left(-\frac{\alpha}{2}\right)\Gamma\left(-\frac{\beta}{2}\right)\Gamma\left(\frac{\alpha+\beta+2\dim}{2}\right)
	} r_{12}^{\alpha + \beta + \dim}\,.
\end{align}
The $\vnxa{1}$ and $\vnxa{2}$ vectors in the integrands of generalized Riesz-type cannot be eliminated via rewriting the vectors into
derivatives. Instead one has to use the averaging procedure in prolate spheroidal coordinates in \cite{Jaranowski:Schafer:1998} 
to get rid of the $\vnxa{}$ vectors. Afterwards one can use the generalized 
Riesz formula which was found by P. Jaranowski during his 3PN point-mass calculations (also given in \cite{Jaranowski:Schafer:1998}), 
\begin{align}
 \left.\int {\rm d}^3 x\, r_1^\alpha r_2^\beta s_{12}^\gamma\right|_{\text{reg}} & = 2\pi \frac{\Gamma(\alpha+2)\Gamma(\beta+2)\Gamma(-\alpha-\beta-\gamma-4)}{\Gamma(-\gamma)} 
[I_{1/2}(\alpha+2,-\alpha-\gamma-2)\nonumber\\ 
& + I_{1/2}(\beta+2,-\beta-\gamma-2)\nonumber\\ 
& - I_{1/2}(\alpha+\beta+4,-\alpha-\beta-\gamma-4) - 1] r_{12}^{\alpha+\beta+\gamma+3}\,,
\end{align}
which reduces to the formula for the integrals of the Riesz-type for $\gamma\to0$. The function $I_{1/2}(x, y)$ is the regularized
incomplete Euler beta function which is defined as
\begin{align}
 I_{1/2}(x, y) & = \frac{B_{1/2}(x,y)}{B(x,y)}\,,
\end{align}
with
\begin{align}
 B_{1/2}(x, y) = \frac{1}{2^x x}\; {}_2 \! F_1\left(1-y,x,x+1;\frac{1}{2}\right)\,, \quad
 B(x,y) = \frac{\Gamma(x)\Gamma(y)}{\Gamma(x+y)}\,,
\end{align}
being the incomplete Euler beta function and the Euler beta function respectively.
Notice that all integrals of the generalized Riesz-type and all inverse Laplacians of two variables can only be solved in $\dim=3$.
Only the UV-singular part of those integrals can be evaluated in $\dim$ dimensions.

It turns out after using the integration procedures mentioned above that all integrands of the generalized Riesz-type appearing
at spin-orbit level have such a structure that the incomplete Euler beta functions appearing there can be reduced to Gamma functions and
Polygamma functions, which could be handled very well by {\scshape Mathematica}.

\section{Result}\label{sec:result}
To check our code we recalculated parts of the 3PN point-mass Hamiltonian given in 
\cite{Jaranowski:Schafer:1998,Damour:Jaranowski:Schafer:2001}. The next-to-next-to-leading order spin-orbit Hamiltonian we obtained as a result of the procedures discussed in Sect. \ref{sec:outline} is given by
\begin{align}
 H^{\text{NNLO}}_{\text{SO}} & = \frac{\gravthree}{\rel^2} \biggl[
	\biggl(
		\frac{7 m_2 (\vmom{1}^2)^2}{16 m_1^5} 
		+ \frac{9 \scpm{\vnun}{\vmom{1}}\scpm{\vnun}{\vmom{2}}\vmom{1}^2}{16 m_1^4} 
		+ \frac{3 \vmom{1}^2 \scpm{\vnun}{\vmom{2}}^2}{4 m_1^3 m_2}\nonumber\\
&		+ \frac{45 \scpm{\vnun}{\vmom{1}}\scpm{\vnun}{\vmom{2}}^3}{16 m_1^2 m_2^2}
		+ \frac{9 \vmom{1}^2 \scpm{\vmom{1}}{\vmom{2}}}{16 m_1^4}
		- \frac{3 \scpm{\vnun}{\vmom{2}}^2 \scpm{\vmom{1}}{\vmom{2}}}{16 m_1^2 m_2^2}\nonumber\\
&		- \frac{3 (\vmom{1}^2) (\vmom{2}^2)}{16 m_1^3 m_2}
		- \frac{15 \scpm{\vnun}{\vmom{1}}\scpm{\vnun}{\vmom{2}} \vmom{2}^2}{16 m_1^2 m_2^2}
		+ \frac{3 \scpm{\vnun}{\vmom{2}}^2 \vmom{2}^2}{4 m_1 m_2^3}\nonumber\\
&		- \frac{3 \scpm{\vmom{1}}{\vmom{2}} \vmom{2}^2}{16 m_1^2 m_2^2}
		- \frac{3 (\vmom{2}^2)^2}{16 m_1 m_2^3}
	\biggr)((\vnun \times \vmom{1})\vspin{1})
	+\biggl(
		- \frac{3 \scpm{\vnun}{\vmom{1}}\scpm{\vnun}{\vmom{2}}\vmom{1}^2}{2 m_1^3 m_2}\nonumber\\
&		- \frac{15 \scpm{\vnun}{\vmom{1}}^2\scpm{\vnun}{\vmom{2}}^2}{4 m_1^2 m_2^2}
		+ \frac{3 \vmom{1}^2 \scpm{\vnun}{\vmom{2}}^2}{4 m_1^2 m_2^2}
		- \frac{\vmom{1}^2 \scpm{\vmom{1}}{\vmom{2}}}{2 m_1^3 m_2}
		+ \frac{\scpm{\vmom{1}}{\vmom{2}}^2}{2 m_1^2 m_2^2}\nonumber\\
&		+ \frac{3 \scpm{\vnun}{\vmom{1}}^2 \vmom{2}^2}{4 m_1^2 m_2^2}
		- \frac{(\vmom{1}^2) (\vmom{2}^2)}{4 m_1^2 m_2^2}
		- \frac{3 \scpm{\vnun}{\vmom{1}}\scpm{\vnun}{\vmom{2}}\vmom{2}^2}{2 m_1 m_2^3}\nonumber\\
&		- \frac{\scpm{\vmom{1}}{\vmom{2}} \vmom{2}^2}{2 m_1 m_2^3}
	\biggr)((\vnun \times \vmom{2})\vspin{1})
	+\biggl(
		- \frac{9 \scpm{\vnun}{\vmom{1}} \vmom{1}^2}{16 m_1^4}
		+ \frac{\vmom{1}^2 \scpm{\vnun}{\vmom{2}}}{m_1^3 m_2}\nonumber\\
&		+ \frac{27 \scpm{\vnun}{\vmom{1}}\scpm{\vnun}{\vmom{2}}^2}{16 m_1^2 m_2^2}
		- \frac{\scpm{\vnun}{\vmom{2}}\scpm{\vmom{1}}{\vmom{2}}}{8 m_1^2 m_2^2}
		\boxed{-\frac{5 \scpm{\vnun}{\vmom{1}} \vmom{2}^2}{16 m_1^2 m_2^2}}\nonumber\\
&		+ \frac{\scpm{\vnun}{\vmom{2}}\vmom{2}^2}{m_1 m_2^3}
	\biggr)((\vmom{1} \times \vmom{2})\vspin{1})
\biggr] \nonumber\\
&+ \frac{\gravthree^2}{\rel^3} \biggl[
	\biggl(
		-\frac{3 m_2 \scpm{\vnun}{\vmom{1}}^2}{2 m_1^2}
		+\left(
			-\frac{3 m_2}{2 m_1^2}
			+\frac{27 m_2^2}{8 m_1^3}
		\right) \vmom{1}^2
		+\left(
			\frac{177}{16 m_1}
			+\frac{11}{m_2}
		\right) \scpm{\vnun}{\vmom{2}}^2\nonumber\\
&		+\left(
			\frac{11}{2 m_1}
			+\frac{9 m_2}{2 m_1^2}
		\right) \scpm{\vnun}{\vmom{1}} \scpm{\vnun}{\vmom{2}}
		+\left(
			\frac{23}{4 m_1}
			+\frac{9 m_2}{2 m_1^2}
		\right) \scpm{\vmom{1}}{\vmom{2}}\nonumber\\
&		-\left(
			\frac{159}{16 m_1}
			+\frac{37}{8 m_2}
		\right) \vmom{2}^2
	\biggr)((\vnun \times \vmom{1})\vspin{1})
	+\biggl(
		\frac{4 \scpm{\vnun}{\vmom{1}}^2}{m_1}
		+\frac{13 \vmom{1}^2}{2 m_1}\nonumber\\
&		+\frac{5 \scpm{\vnun}{\vmom{2}}^2}{m_2}
		+\frac{53 \vmom{2}^2}{8 m_2}
		- \left(
			\frac{211}{8 m_1}
			+\frac{22}{m_2}
		\right) \scpm{\vnun}{\vmom{1}} \scpm{\vnun}{\vmom{2}}\nonumber\\
&		-\left(
			\frac{47}{8 m_1}
			+\frac{5}{m_2}
		\right)\scpm{\vmom{1}}{\vmom{2}}
	\biggr)((\vnun \times \vmom{2})\vspin{1})
	+\biggl(
		-\left(
			\frac{8}{m_1}
			+\frac{9 m_2}{2 m_1^2}
		\right)\scpm{\vnun}{\vmom{1}}\nonumber\\
&		+\left(
			\frac{59}{4 m_1}
			+\frac{27}{2 m_2}
		\right)\scpm{\vnun}{\vmom{2}}
	\biggr)((\vmom{1} \times \vmom{2})\vspin{1})
\biggr]\nonumber\\
&+\frac{\gravthree^3}{\rel^4} \biggl[
	\left(
		\frac{181 m_1 m_2}{16}
		+ \frac{95 m_2^2}{4}
		+ \frac{75 m_2^3}{8 m_1}
	\right) ((\vnun \times \vmom{1})\vspin{1})\nonumber\\
&	- \left(
		\frac{21 m_1^2}{2}
		+ \frac{473 m_1 m_2}{16}
		+ \frac{63 m_2^2}{4}
	\right)((\vnun \times \vmom{2})\vspin{1})
\biggr]
 + (1\leftrightarrow2)\,.
\end{align}
Obviously there are no logarithmic dependencies of $\rel$ appearing.\footnote{The
published version of the present article contains a typo in the framed term,
the coefficient should read $-\frac{5}{16}$ instead of $-\frac{15}{16}$.
%The result given here now is correct.
We thank S. Marsat for pointing this out.
Notice that the term in question does not contribute in the center-of-mass frame.}
Also notice that the canonical antisymmetric spin tensor was rewritten
in terms of the canonical spin vector, which is possible in $\dim=3$.
The $\dim$-dimensional UV-analysis described in 
\cite{Damour:Jaranowski:Schafer:2001, Damour:Jaranowski:Schafer:2008:2}
and Sect. \ref{sec:outline} gave contributions to intermediate expressions, however
they exactly canceled in the final result. In contrast, for point-masses only 
the poles in $\varepsilon=\dim-3$ canceled but a finite part remained.
Further from a combinatorial point of view there are 66 algebraically 
different possible contributions to the Hamiltonian for each object 
(written in terms of the canonical spin tensor), but 24 of them do not 
appear in the canonical representation used here.
The Hamiltonian is valid for any compact objects like black holes or neutron stars. It completes 
the knowledge of the dynamics up to and including 3.5PN for maximally rotating black holes. 
For other objects the $S^3$ Hamiltonians and the inclusion of tidal effects are missing. 
One can find a discussion of leading order tidal effects in \cite{Damour:Nagar:2009}.
Notice that the coupling structure in this Hamiltonian reduces in the center-of-mass frame 
to a pure $\scpm{\vct{L}}{\vct{S}}$ structure with complicated coefficients. So the Hamiltonian 
is indeed a spin-orbit Hamiltonian. We compared our result in the test-spin limit with the PN 
expanded Hamiltonians of a test-spin near a Kerr black hole in ADM coordinates in 
\cite[Eq. (6.20)]{Barausse:Racine:Buonanno:2009} and got full agreement.\footnote{Note
that there is a typo in \cite{Barausse:Racine:Buonanno:2009} in Eq. (6.20), the last term 
$\frac{105}{8}\frac{M^2}{r^5}(\vct{S}^{\ast}\cdot\vct{L})$ has to be $\frac{75}{8}\frac{M^2}{r^5}(\vct{S}^{\ast}\cdot\vct{L})$.
See the arXiv version of \cite{Barausse:Racine:Buonanno:2009} for the correct equation. 
We thank E.\ Barausse and A.\ Buonanno for clarifying this issue.}

The matter variables appearing in the fully reduced matter-only Hamiltonian fulfill the standard Poisson bracket relations, namely
\begin{align}
 \{\hat{z}^i_a, \mom{a}{j}\} = \delta_{ij}\,, \quad
 \{\hat{S}_{a\,(i)}, \hat{S}_{a\,(j)}\} = \varepsilon_{ijk}\hat{S}_{a\,(k)}\,,
\end{align}
all other zero and the Hamiltonian can be used to get the time evolution of an arbitrary phase space 
function $A$ via
\begin{align}
 \frac{\text{d}A}{\text{d}t} &= \{A,H\} + \frac{\partial A}{\partial t}\,.
\end{align}

Although the algorithms to be used at formal 3PN level are given above, the whole calculation 
is very hard. It is in particular much harder than (and really different from) the formal 2PN calculation
at next-to-leading order spin-orbit level.

\section{Approximate Poincar\'e algebra}\label{sec:poincare}
In this section we check that the global Poincar\'e algebra is fulfilled in a PN approximate way,
see, e.g., \cite{Damour:Jaranowski:Schafer:2000, Damour:Jaranowski:Schafer:2008:1}.
Besides the Hamiltonian, the quantities entering the Poincar\'e algebra are
the center-of-mass vector $\vct{G}$, the total linear momentum $\vct{P}$, and
the total angular momentum tensor $J^{ij} = - J^{ji}$. As the latter two are the
infinitesimal generators of translations and rotations, they can be expressed in
terms of canonical variables as \cite{Damour:Jaranowski:Schafer:2000, Damour:Jaranowski:Schafer:2008:1}
\begin{align}
 \vct{P} = \sum_a \vmom{a}\,, \quad
 J^{ij} = \sum_a \left[\hat{z}^i_a \mom{a}{j} - \hat{z}^j_a \mom{a}{i} + \spin{a}{i}{j}\right]\,,
\end{align}
where the canonical spin tensor $\spin{a}{i}{j}$ is related to the canonical spin vector $\hat{\vct{S}}_{a}$ via
$\spin{a}{i}{j} = \varepsilon_{ijk} \hat{S}_{a\,(k)}$.
Notice that the total angular momentum is not only the sum of the orbital angular 
momenta but also contains the spin angular momenta. For the contributions of
the propagating field degrees of freedom see, e.g., \cite{Steinhoff:Wang:2009,Steinhoff:2011}.
As in \cite{Damour:Jaranowski:Schafer:2000, Damour:Jaranowski:Schafer:2008:1} we
used an ansatz for the center-of-mass vector $\vct{G}$ at next-to-next-to-leading order spin-orbit level which 
contains 68 unknown coefficients here. For comparison we mention that the 2PN binary point-mass $\vct{G}$-vector requires 20 unknown coefficients
to be fixed and the 3PN binary point-mass $\vct{G}$-vector requires 78 unknown coefficients to be fixed 
\cite{Damour:Jaranowski:Schafer:2000}. In \cite[Eq. (5.9)]{Damour:Jaranowski:Schafer:2008:1} one can see the ansatz 
for the next-to-leading order case (which contains only 8 unknown coefficients). But at the order considered here 
there will be additional linear momentum powers which increase the number of necessary coefficients significantly.
16 of them can be fixed by taking into account the $\{G^i, P^j\}$ Poisson bracket relation
appearing in the Poincar\'e algebra.
The remaining 52 coefficients were uniquely fixed by evaluating the $\{\vct{G}, H\}$ Poisson 
brackets. The consistency of the solution obtained by evaluating the Poisson bracket
above was checked by evaluating the $\{G^i, G^j\}$ Poisson bracket relation and all
remaining relations of the Poincar\'e algebra.

The center-of-mass vector at next-to-next-to-leading order spin-orbit level is given by
\begin{align}
  \vct{G}^{\text{NNLO}}_{\text{SO}} & = 
 \frac{(\vmom{1}^2)^2}{16 m_1^5} (\vmom{1}\times\vspin{1})\nonumber\\
&+ (\vmom{2}\times\vspin{1})
 	\biggl[
 		\frac{\gravthree}{\rel}
 			\biggl(
 				- \frac{3 \scpm{\vnun}{\vmom{1}}\scpm{\vnun}{\vmom{2}}}{8 m_1 m_2}
				- \frac{\scpm{\vmom{1}}{\vmom{2}}}{8 m_1 m_2}
			\biggr)
		+\frac{\gravthree^2}{\rel^2} 
			\biggl(
				-\frac{47 m_1}{16} 
				-\frac{21 m_2}{8}
			\biggr)
	\biggr]\nonumber\\
&+ (\vmom{1}\times\vspin{1})
	\biggl[
		\frac{\gravthree}{\rel}
			\biggl(
				\frac{9 m_2 \vmom{1}^2}{16 m_1^3}
				- \frac{5 \vmom{2}^2}{8 m_1 m_2}
			\biggr)
		+\frac{\gravthree^2}{\rel^2} 
			\biggl(
				\frac{57 m_2}{16} 
				+\frac{15 m_2^2}{8 m_1}
			\biggr)
	\biggr]\nonumber\\
&+ (\vnun\times\vspin{1})
	\biggl[
		\frac{\gravthree}{\rel}
			\biggl(
				\frac{9\scpm{\vnun}{\vmom{1}}\scpm{\vnun}{\vmom{2}}^2}{16 m_1 m_2}
				+ \frac{\scpm{\vnun}{\vmom{2}}\scpm{\vmom{1}}{\vmom{2}}}{8 m_1 m_2}
				+ \frac{\scpm{\vnun}{\vmom{1}}\vmom{2}^2}{16 m_1 m_2}
			\biggr)\nonumber\\
&\quad		+\frac{\gravthree^2}{\rel^2}
			\biggl(
				-\frac{5 m_2}{8} \scpm{\vnun}{\vmom{1}}
				+\left\{\frac{13 m_1}{8} + \frac{11 m_2}{4}\right\} \scpm{\vnun}{\vmom{2}}
			\biggr)
	\biggr]\nonumber\\
&- \frac{\gravthree}{\rel} \vmom{1} \frac{\scpm{\vnun}{\vmom{2}}((\vnun \times \vmom{2})\vspin{1})}{2 m_1 m_2}\nonumber\\
&+ \frac{\gravthree}{\rel} \vmom{2}
	\biggl(
		-\frac{\scpm{\vnun}{\vmom{2}}((\vnun \times \vmom{1})\vspin{1})}{8 m_1 m_2}
		+\frac{\scpm{\vnun}{\vmom{1}}((\vnun \times \vmom{2})\vspin{1})}{2 m_1 m_2}\nonumber\\
&\quad		-\frac{((\vmom{1} \times \vmom{2})\vspin{1})}{8 m_1 m_2}
	\biggr)\nonumber\\
&+ \vnun
	\biggl[
		\frac{\gravthree}{\rel}
			\biggl(
				\biggl\{
					\frac{m_2 \vmom{1}^2}{16 m_1^3}
					+ \frac{15 \scpm{\vnun}{\vmom{2}}^2}{16 m_1 m_2}
					- \frac{3 \vmom{2}^2}{16 m_1 m_2}
				\biggr\}((\vnun \times \vmom{1})\vspin{1})\nonumber\\
&\quad\quad				+\biggl\{
					-\frac{3 \scpm{\vnun}{\vmom{1}}\scpm{\vnun}{\vmom{2}}}{2 m_1 m_2}
					-\frac{\scpm{\vmom{1}}{\vmom{2}}}{2 m_1 m_2}
				\biggr\}((\vnun \times \vmom{2})\vspin{1})\nonumber\\
&\quad\quad			+\frac{13 \scpm{\vnun}{\vmom{2}}}{8 m_1 m_2}((\vmom{1} \times \vmom{2})\vspin{1})
			\biggr)\nonumber\\
&\quad		+\frac{\gravthree^2}{\rel^2}
			\biggl(
				\left\{\frac{m_2}{2} + \frac{5 m_2^2}{4 m_1}\right\} ((\vnun \times \vmom{1})\vspin{1})
				+\left\{-2 m_1 - 5 m_2\right\} ((\vnun \times \vmom{2})\vspin{1})
			\biggr)
	\biggr]\nonumber\\
&+ \frac{\hat{\vct{z}}_1}{\rel}
	\biggl[
		\frac{\gravthree}{\rel}
			\biggl(
				\biggl\{
					\frac{3 \scpm{\vnun}{\vmom{1}} \scpm{\vnun}{\vmom{2}}}{m_1 m_2}
					+\frac{\scpm{\vmom{1}}{\vmom{2}}}{m_1 m_2}
				\biggr\} ((\vnun \times \vmom{2})\vspin{1})\nonumber\\
&\quad\quad			+\biggl\{
					\frac{3 \scpm{\vnun}{\vmom{1}}}{4 m_1^2}
					-\frac{2 \scpm{\vnun}{\vmom{2}}}{m_1 m_2}
				\biggr\}((\vmom{1} \times \vmom{2})\vspin{1})\nonumber\\
&\quad\quad			+\biggl\{
					- \frac{5 m_2 \vmom{1}^2}{8 m_1^3}
					- \frac{3 \scpm{\vnun}{\vmom{1}} \scpm{\vnun}{\vmom{2}}}{4 m_1^2}
					- \frac{3 \scpm{\vnun}{\vmom{2}}^2}{2 m_1 m_2}\nonumber\\
&\quad\quad\quad			- \frac{3 \scpm{\vmom{1}}{\vmom{2}}}{4 m_1^2}
					+ \frac{3 \vmom{2}^2}{4 m_1 m_2}
				\biggr\} ((\vnun \times \vmom{1})\vspin{1})
			\biggr)\nonumber\\
&		+\frac{\gravthree^2}{\rel^2}
			\biggl(
				\left\{
					-\frac{11 m_2}{2}
					-\frac{5 m_2^2}{m_1}
				\right\} ((\vnun \times \vmom{1})\vspin{1})
				+\left\{
					6 m_1
					+ \frac{15 m_2}{2}
				\right\} ((\vnun \times \vmom{2})\vspin{2})
			\biggr)
	\biggr]\nonumber\\
& + (1\leftrightarrow2)\,.
\end{align}
%
% document ends here
%

\ifnotadp
\paragraph*{Acknowledgments}
\ack
\setlength{\bibsep}{0pt}
\bibliographystyle{utphys}
\bibliography{../references}
% generate 3PNSOPaperRapid.bib from master database references.bib:
%	jabref -n -a 3PNSOPaperRapid.aux,3PNSOPaperRapid.bib ../references.bib
%\bibliography{3PNSOPaperRapid}
%\input{3PNSOPaperRapid_refs}
\fi

\ifadp
\begin{acknowledgement}
\ack
\end{acknowledgement}
\input{refs_adp}
\fi

\end{document}

%% file: macros.tex
\def\vct#1{\mathbf{#1}}

\def\nunit{n}%_{\rm{12}}}

\newcommand{\vnunit}{\vct{\nunit}_{12}}

\newcommand{\spin}[3]{\hat{S}_{#1\, (#2)(#3)}}
\def\canmom{P}
\newcommand{\mom}[2]{{\canmom}_{#1\,#2}}

\newcommand{\vmom}[1]{{\vct{\canmom}}_{#1}}
\newcommand{\vnxa}[1]{{\vct{n}}_{#1}}
\newcommand{\vnun}{{\vnunit}}
\newcommand{\dl}[1]{\delta_{#1}}
\newcommand{\scpm}[2]{\left(#1\,#2\right)}
\newcommand{\vspin}[1]{\hat{\vct{S}}_{1}}

\def\gravthree{G}
\def\dim{d}
\def\rel{r_{12}} % subst r and r_{12} by \rel